# Photorefractive and pyroelectric photonic memory and long-term stability in thin-film lithium niobate microresonators

*Xinyi Ren[1,2], Chun-Ho Lee[1,2], Kaiwen Xue[1], Shaoyuan Ou[1], Yue Yu[1], Zaijun Chen[1], and Mengjie Yu[1]\**

[1]Ming Hsieh Department of Electrical and Computer Engineering, University of Southern California, Los Angeles, California 90089, USA
[2]These authors contributed equally: Xinyi Ren, Chun-Ho Lee.
E-mail: mengjiey@usc.edu

The stability of the integrated photonic circuits is of critical importance for many applications that require high frequency precision or robust operation over time, such as optomechanical sensing, frequency conversion, optical communication, and quantum optics. Photonic memory is useful for low-energy optical computing and interconnects. Thin film lithium niobate (TFLN), as an emerging photonic platform, exhibits complex material properties including pyroelectric (PE) and photorefractive (PR) effects which could lead to intra-device drift and excess noise under different environmental or operating conditions as well as be utilized for building photonic memory. However, the long-term stability and memory effect of its optical properties has not been explored. In this paper, we discovered a long-lived change of optical refractive index as a result of light excitation and temporal temperature variation using Z-cut TFLN microresonators and reveal a strong dependence of the instability with the crystal orientation of the thin film form. The recovery time are measured to be over 10 hours. Leveraging the photonic memory with a long relaxation time, we realize optical trimming of the cavity resonance frequencies. Our result offers insights towards understanding the fundamental noise properties and dynamic behavior of the integrated TFLN material and devices.

## 1. Introduction

Lithium niobate (LN), first discovered in 1949 as a ferroelectric crystal, has been known for its excellent optical properties including such as large electric-optic (EO) coefficient, second and third-order optical nonlinearities, and a wide transparency window from 0.3 to 5.5 μm[1-3]. The recent advancement of the low-loss thin-film-lithium-niobate (TFLN) photonic devices has enabled compact high-performance frequency converters[4,5], EO[6,7] and acousto-optic modulators[8] on a monolithic platform, which quickly attract significant interest for building the future integrated photonic circuits. However, the scalability and reliability of the TFLN platform remains an open but critically important question with many technical challenges to be solved, such as intra-device drift, inter-device variation and fabrication inhomogeneities.

**Figure 1a** shows that the temporal variation of the optical refractive index would lead to the drift of the bias voltage in the EO modulators which degrades the health of the classical and quantum optical communication links[9,10] as well as the drift of the cavity resonance position which disrupts the measurement required at a certain pump-cavity detuning[11]. Despite the studies of the bulk LN material[12], the TFLN platform shows different behaviors from its bulk form due to the tighter optical confinement, the condition of surface defects and nanofabrication process[13,14]. Compared to the more matured photonic material, such as silicon or silicon nitride, the LN exhibits far richer material properties, such as birefringent, photorefractive (PR), photovoltaic and pyroelectric (PE) effect. Photorefractive effect is a result of the photovoltaic effect through optically pumping the trapped electrons from the impurity level to the conduction band (**Fig. 1b**) while pyroelectric effect describes the change of the spontaneous polarization of the material as a result of temperature change $\Delta T$ (**Fig. 1c**), both of which give rise to a built-in electrical field along the crystal axis as sketched in **Fig. 1d**. The change of the optical refractive index is proportional to the product of the EO coefficient and the resulting electric field. Therefore, experimental characterization of the PR and PE effects is essential to understand the systems which need to operate at high optical power[15] or be cooled to cryogenic temperature[10], rely on thermo-optic effect for tunability[16-18] and are sensitive to the carrier dynamics such as superconducting detectors[19,20]. There are recent reports on the fast time oscillation of the PR effect[13,21,22] and the thermo-refractive noise[23]. However, the long-term stability of the optical properties has not been studied on the TFLN platform. In addition, non-volatile photonic memory[24] has not been reported in TFLN devices despite the photorefractive effect is routinely used for holographic data storage in bulk LN crystals[25,26].

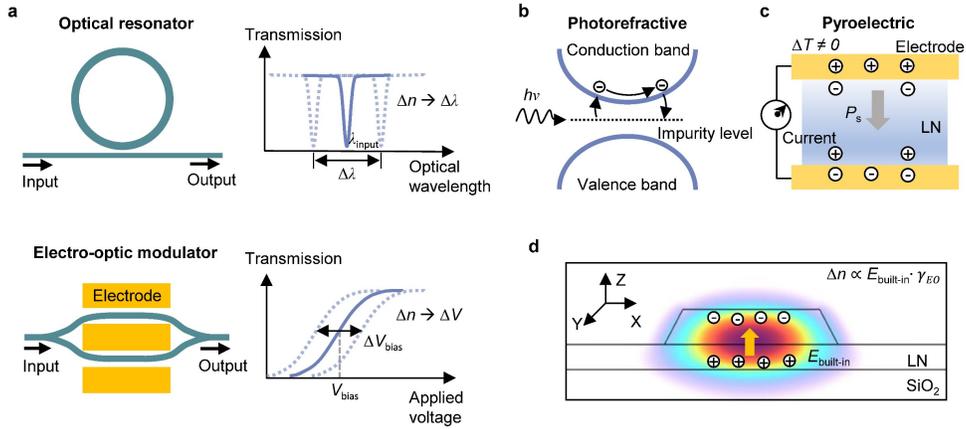

**Fig. 1**| Stability of the thin-film lithium niobate photonics. **a** The change of the optical refractive index ($\Delta n$) could lead to the transmission variation due to the resonance frequency drift of an optical cavity or the bias-voltage drift of an amplitude electro-optic (EO) modulator. **b** Electrons from the impurity level is optically excited into the conduction band by the photorefractive (PR) effect; **c** Temperature change induces electrical current by the pyroelectric (PE) effect, $P_s$: spontaneous polarization of the material; **d** Carriers from either the PR or PR effect lead to a built-in electrical field ($E_{\text{built-in}}$) in the TFLN waveguide which results in $\Delta n$ via the EO effect of LN ($\Delta n \propto E_{\text{built-in}} \cdot \gamma_{\text{EO}}$, $\gamma_{\text{EO}}$: the EO coefficient).

In this paper, we systematically investigate the PR and PE effects and their temporal dynamics in both X-cut and Z-cut TFLN devices. The change of the optical refractive index is monitored using the transmission spectrum of high quality-factor microresonators. Due to the electron trapping at the surface defect states and the thin feature of the film, the built-in electric field is found to be stronger and takes significantly longer time to decay when the crystal axis is perpendicular to the device plane (Z-cut), in comparison with the X-cut device with a similar dimension. In Z-cut TFLN microresonators, we observe a blue-shifted cavity resonance with a recovery time of over 20 hours after optical excitation is turned off. Leveraging the long lifetime, we achieve deterministic and controllable frequency shifting of the resonance frequency via controlling the injected light, and trimming different devices to resonant at identical optical frequencies. Moreover, we find that the joint PE and EO effect causes the optical resonance to blue shift upon increasing the device temperature. The time scale of the process is dominated by the charge dynamics and differs significantly from the thermal dynamics. Similarly, a long recovery time of over 10 hours is observed. Our work is of importance to understand the stability of the TFLN photonic devices for both classical and quantum applications. The long-lived PR and PE effect could also be used for realizing non-volatile optical memory and overcoming fabrication inhomogeneity as well as for optical sensors[27] and bolometers[28].

## 2. Results

### 2.1. Approach

The devices are air cladded and fabricated on a 600-nm LN wafer with an etch depth of 350 nm, a top width of 1.5 μm and a bending radius of 80 μm. The cross-section and scanning electron microscope images are shown in **Fig 2a**. **Figure 2b** shows the transmission spectrum of the microresonator via sweeping the frequency of a continuous-wave laser. The intrinsic $Q$ factor is extracted to be $3.3 \times 10^6$ at near 1560 nm for the fundamental transverse-electric (TE) mode. Both X-cut and Z-cut LN microrings with the same geometry are fabricated for comparative analysis. The refractive index change ($\Delta n$) can be extracted by measuring the optical resonance shift of microrings based on

$$\Delta n = n \cdot \frac{\Delta \lambda}{\lambda} \tag{1}$$

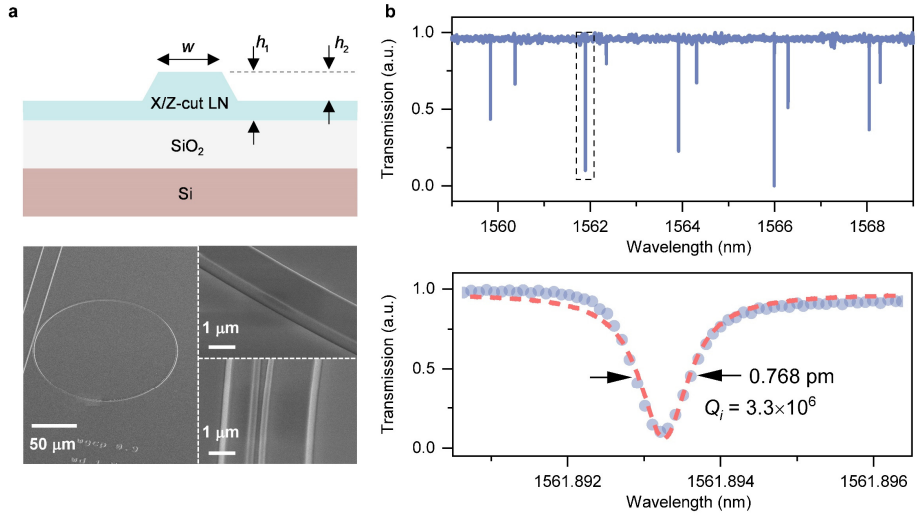

**Fig. 2**| Characterization of the TFLN devices. **a** Cross-section schematic and the scanning electron microscope (SEM) of the microresonators. The waveguide height $h_1$, 600 nm; the etch depth $h_2$, 350 nm; the waveguide top width $w$, 1.5 μm. **b** Transmission spectrum. The extracted intrinsic Q factor ($Q_i$) to be 3.3 million.

### 2.2. Photorefractive effect characterization

To study the photorefractive induced refractive index change, we control the sweeping parameter of a continuous-wave control laser across the resonance, and then measure the optical resonance shift using a second probe laser at low optical power, as shown in **Fig. 3a**. The device is placed on a thermo-electric-controller (TEC) to keep a constant temperature. The optical power of the control laser in the waveguide is 80 mW. The detailed setup is shown in **Figure S1** in the supplementary file. After 500 cycles of the control laser sweeping at a 5-nm/s speed across a 20-pm spectral range, a total cumulated wavelength shift of -13.7 pm is achieved via the PR effect (**Fig. 3b**) corresponding to a $\Delta n$ of $-1.8\times10^{-5}$. The index is preserved after the control light is turned off, indicating a photonic memory effect which has not been reported before.

**Figure 3c** plots the resonance shift (and $\Delta n$) is linearly proportional to the number of the sweeping cycle and the control laser power, thus the total injected optical energy. The largest refractive index change we demonstrate in this work is $-8.6\times10^{-5}$. One can estimate the internal electrical field strength $E_{\text{built-in}}$ along the polarization axis to be $1.7\times10^{6}$ V/m based on

$$\Delta n = -\frac{1}{2}n_o^3 \cdot r_{13} \cdot E_{\text{built-in}} \quad (2)$$

where $n_0$ and $r_{13}$ are the LN ordinary optical refractive index and the EO coefficient (9.6 pm/V)[29], respectively. We also verify that the optical loss of the device didn't deteriorate by monitoring the $Q$-factor during the process (**Fig. 3d**). Furthermore, we demonstrate the PR-based tuning for two different rings, both of which are trimmed to have the same resonance wavelength of 1565.85 nm by a blue shift of 53 pm ($\Delta n = -6.7\times10^{-5}$) and 17 pm ($\Delta n = -2.2\times10^{-5}$), respectively (**Fig. 4a**). The non-volatile tuning shows promises for post-fabrication trimming of photonic circuits. Next, we study the temporal stability of the PR-induced effect. As shown in **Fig. 4b**, the cavity resonance shows a change of < 1 pm (comparable with the loaded linewidth) over a 10-min time period. As we continue to track the shift over a time window of 20 hours, we observe that the resonance position slowly moves towards the original wavelength due to the space-charge field relaxation. After 20 hours, the optical index doesn't restore to the initial value. This process can be described by

$$\Delta\lambda = \partial_1 \cdot e^{-\tau_1/t} + \partial_2 \cdot e^{-\tau_2/t} \quad (3)$$

where $t$ is the time, $\partial_1$ and $\partial_2$ denote the wavelength shift before recovery, $\tau_1$ and $\tau_2$ are characteristic time constants. The time constants $\tau_1$ and $\tau_2$ are fitted to be 1.69 and 10.06 hours respectively, five orders of magnitude longer than the reported values in the X-cut LN[13,21,22]. The detailed analysis and comparison are shown in the discussion section. The large time constant could be leveraged to realize temporary optical memory on chip.

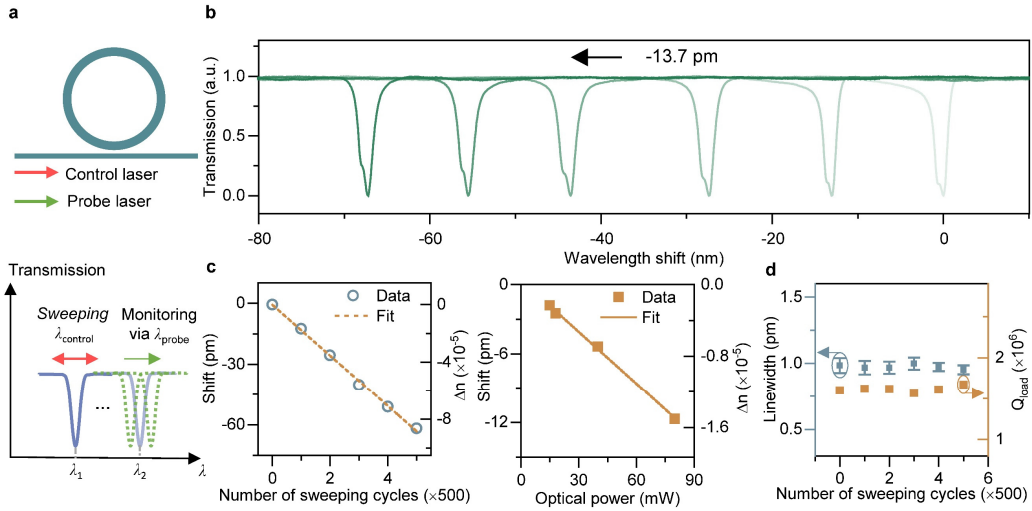

**Fig. 3|** Photorefractive effect characterization. **a** Schematic of the PR characterization. A control laser is used to sweep the resonance, after which the probe laser is to monitor the resonance shift. **b** Resonance blueshifts during 2500 sweeping loops, each 500 cycles induce -13.7 pm wavelength shift. **c** Resonance wavelength shift ($\Delta n$) as a linear function of the number of sweeping cycles (at 80-mW on-chip optical power) and the on-chip power (with 430 cycles). **d** Loaded linewidth and the loaded Q factor during the 2500 sweeping cycles.

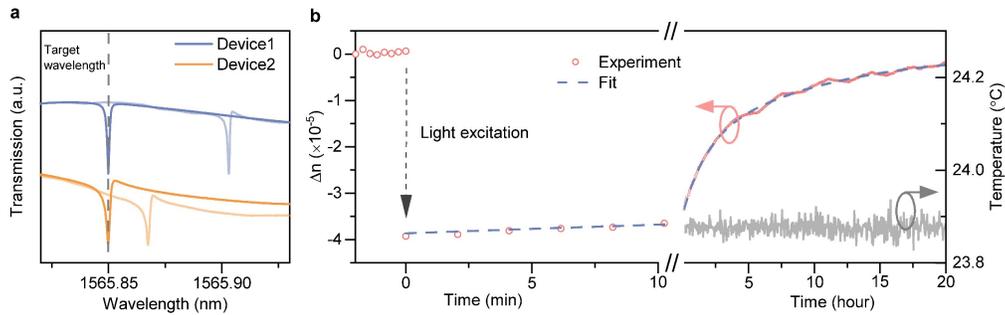

Fig 4| Optical trimming and stability measurement. **a** Optical resonance trimming via the PR effect. **b** Temporal dynamics of the optical refractive index. PR effect shows a relaxation process over a duration of 20 h with a change of less than 1 pm within the first 10 min window.

## 2.3. Pyroelectric effect characterization

We characterize the pyroelectric effect by changing the device temperature using the TEC while tracking the resonance position using the same probe laser. The temperature variation $\Delta T$ modifies the position of the atoms within the crystal and thus changes the polarization of the LN material, generating a voltage across the thin film LN via the PE effect. The resulting electrical field is given by

$$E_{\text{built-in}} = -\frac{1}{\varepsilon_0 \varepsilon_r} \cdot \frac{\partial P_s}{\partial T} \cdot \Delta T = -\frac{p \cdot \Delta T}{\varepsilon_0 \varepsilon_r} \quad (4)$$

where $p_s$ is the spontaneous polarization, $p = \frac{\partial P_s}{\partial T}$ represents the pyroelectric coefficient of LN ($C \cdot cm^2 \cdot °C^{-1}$), $\varepsilon_0$ and $\varepsilon_r$ denote the free space permittivity and relative dielectric permittivity, respectively. Therefore, the $\Delta n$ via the PE and EO effect can be estimated using equation (2), indicating a negative relationship between $\Delta n$ and $\Delta T$. We note that the thermo-optic (TO) effect would take place concurrently and contribute to a positive change of the index with an increased temperature. ($\Delta n = dn/dT \cdot \Delta T$, where $dn/dT$ is the TO coefficient[30,31]).

**Figure 5a** plots the resonance shift and the index change as a function of time at a temporal resolution of 3s over a 30-min window, while we apply a temporal pulse of temperature with a peak temperature of 54°C. We observe a dynamic resonance shift with four stages as labelled in **Fig. 5a**: 1) the resonance blue shifts proportional to the temperature difference ($\Delta T > 0$) due to the PE and EO effect; 2) the resonance starts to red shift due to the slower TO effect; 3) the resonance experiences a sharp transition of red shift due to the negative $\Delta T$ ($< 0$) and the PE; 4) resonance blue shifts due to the TO effect as the temperature goes back to the room temperature, and continue to blue shift due to the slow relaxation of PE-induced carriers. It is clear that the TO and PE are distinct from each other in terms of shifts and the time scale, as the TO effect is determined by thermal dissipation while the PE effect is related to the carrier relaxation dynamics (likely due to the surface charge neutralization through the air). For clarity, we show the corresponding transmission spectra for these stages in **Fig. 5b**.

**Figure 5c** shows the temporal dynamics under pulses with different peak temperature values of 54, 63, 66 and 70 °C, respectively. We extract the PE-induced $dn/dT$ to be $-2.9 \times 10^{-6}/°C$. In the final (4$^{th}$) stage, there are two different decay times, one of which is relatively fast attributed to the TO effect (shaded in yellow). The slow decay (shaded in gray) is due to the PE effect. **Fig. 5d** shows the recovery time for the PE effect to be over 10 hours. The fitted time constants are shown in Table S1. This result presents the first characterization of the PE effect and its long-term stability in the TFLN platform. It is of importance for applications including chip-based temperature sensor and bolometer.

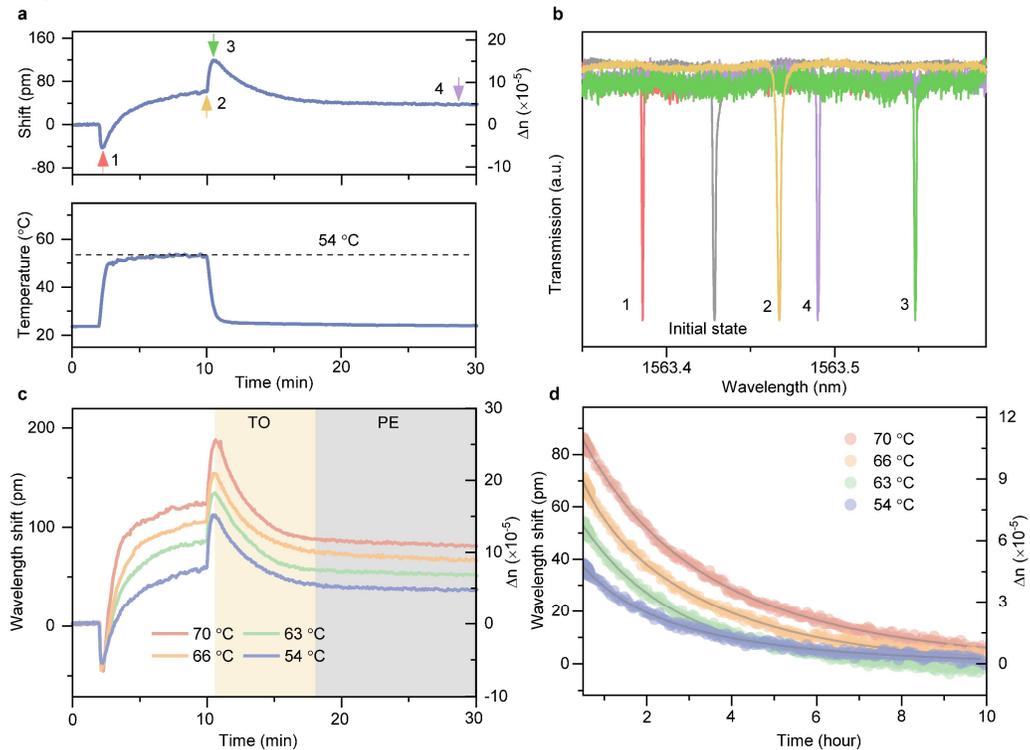

**Fig. 5|.** Pyroelectric effect characterization. **a** Resonance shifts over 30 mins under excitation of a temperature pulse. **b** The transmission spectra of four stages labelled in (a). **c** Temporal dynamics with four different temperature pulses. The two shaded regimes represent processes dominated by the TO and PE effects, respectively. **d** PE relaxation process over 10 hours.

## 3. Discussion and conclusion

Finally, we perform similar investigation in the X-cut microresonators of the same dimension. In the PR experiment, the long-term relaxation and the memory phenomenon is no longer observable in the X-cut device and the optical resonance goes back to the initial position after the control laser sweep. Instead, we observe the distortion of the optical resonance transmission as we bidirectionally scan across the resonance (**Fig. S2**). This is a result of the PR effect at a faster time scale, which competes with the TO effect [13,21,22]. Similarly, the long-term instability of the PE effect is also absent in X-cut devices and the refractive index change follows the thermal dynamics (**Fig. S3**). Therefore, the memory effect strongly depends on the crystal orientation and poses challenges deploying the Z-cut TFLN devices for precision optical measurement, suggesting that active feedback circuits might be necessary. Moreover, it is worthy to note that the resulting internal electric field is stronger as one further shrinks the physical dimension of the device (tighter optical confinement), which explains distinct behaviors from its bulk form. The slow recombination of the electrons at the surface could explain the long relaxation time of optical refractive index observed on the Z-cut devices. We envision that applying conductive coating to neutralize the surface charge[12,14] could further reduce the recovery time or erase the optical memory. Furthermore, the induced electrical field should be of relevance to the optical poling of the LN[32] as well as the acoustic circuit operation via piezoelectricity[33].

In summary, we demonstrate a temporary photonic memory effect induced by light and temperature excitation through the photorefractive, pyroelectric and electro-optic effects based on the high-$Q$ TFLN microresonators. Long-lived optical refractive index change over hours is observed and characterized in the Z-cut TFLN, showing distinct behaviors from the X-cut TFLN and revealing the temporal dynamics of the charge dissipation. Taking advantage of the long-lived feature, non-volatile trimming of TFLN microresonators is demonstrated, for the first time, via a precise control of light excitation. Our work is an important step towards building a practical, scalable and reliable TFLN photonic platform in the future.

## Acknowledgements

This work is supported by the startup funding provided by Ming Hsieh Department of Electrical and Computer Engineering at Universtity of Southern California. The device fabrication was performed at the John O'Brien Nanofabrication Laboratory. The views, opinions and/or findings expressed are those of the authors and should not be interpreted as representing the official views or policies of the Department of Defense or the US government.

## Author contributions

M.Y. conceived the idea. C.L. designed the chip and fabricated the devices. X.R. designed the experiment and carried out the measurements. X.R., C.L. and M.Y. analyzed the data with the help of K.X., S.O., Y.Y and Z.C.. X.R. drafted the manuscript with contribution from all authors. M.Y. revised the manuscript. M.Y. and Z.C. supervised the project.

## Competing interests

The authors declare no competing interests.

## Data availability

The data that support the findings of this study are available from the corresponding author upon reasonable request.